# Superconductivity and orbital-selective nematic order in a new titanium-based kagome metal CsTi$_3$Bi$_5$

Haitao Yang[1,2,3#], Yuhan Ye[1,2#], Zhen Zhao[1,2#], Jiali Liu[1,2#], Xin-Wei Yi[2,3#], Yuhang Zhang[1,2], Jinan Shi[2], Jing-Yang You[2,3], Zihao Huang[1,2], Bingjie Wang[1,2], Jing Wang[1,2], Hui Guo[1,2,3], Xiao Lin[2,3], Chengmin Shen[1,2], Wu Zhou[2,3], Hui Chen[1,2,3\*], Xiaoli Dong[1,2,3\*], Gang Su[2,3\*], Ziqiang Wang[4\*] and Hong-Jun Gao[1,2,3\*]

[1] Beijing National Center for Condensed Matter Physics and Institute of Physics, Chinese Academy of Sciences, Beijing 100190, PR China

[2] School of Physical Sciences, University of Chinese Academy of Sciences, Beijing 100190, PR China

[3] CAS Center for Excellence in Topological Quantum Computation, University of Chinese Academy of Sciences, Beijing 100190, PR China

[4] Department of Physics, Boston College, Chestnut Hill, MA 02467, USA

[#]These authors contributed equally to this work

[*]Correspondence to: hjgao@iphy.ac.cn, wangzi@bc.edu, gsu@ucas.ac.cn, dong@iphy.ac.cn, hchenn04@iphy.ac.cn

**Fabrication of new types of superconductors with novel physical properties has always been a major thread in the research of superconducting materials. An example is the enormous interests generated by the cascade of correlated topological quantum states in the newly discovered vanadium-based kagome superconductors $AV_3Sb_5$ (A=K, Rb, and Cs) with a $Z_2$ topological band structure[1-13]. Here we report the successful fabrication of single-crystals of titanium-based kagome metal $CsTi_3Bi_5$ and the observation of superconductivity and electronic nematicity. The onset of the superconducting transition temperature $T_c$ is around 4.8 K. In sharp contrast to the charge density wave superconductor $AV_3Sb_5$, we find that the kagome superconductor $CsTi_3Bi_5$ preserves translation symmetry, but breaks rotational symmetry and exhibits an electronic nematicity. The angular-dependent magnetoresistivity shows a remarkable two-fold rotational symmetry as the magnetic field rotates in the kagome plane. The scanning tunneling microscopy and spectroscopic imaging detect rotational-symmetry breaking $C_2$ quasiparticle interference patterns (QPI) at low energies, providing further microscopic evidence for electronic nematicity. Combined with first-principle calculations, we find that the nematic QPI is orbital-selective and dominated by the Ti $d_{xz}$ and $d_{yz}$ orbitals, possibly originating from the intriguing orbital bond nematic order. Our findings in the new "135" material $CsTi_3Bi_5$ provide new directions for exploring the multi-orbital correlation effect and the role of orbital or bond order in the electron liquid crystal phases evidenced by the symmetry breaking states in kagome superconductors.**

CsTi$_3$Bi$_5$ (CTB) has a hexagonal crystal structure in the space group of P6/mmm (Fig. 1**a**), which is isostructural to the vanadium-based kagome metal $A$V$_3$Sb$_5$. The compound forms a layered structure. The perfect kagome layer of the Ti atoms is coordinated by a net of Bi1 atoms located at the center of the hexagons, which is sandwiched between two additional honeycomb layers of Bi2 atoms on top and bottom of the Ti triangles in the kagome plane. The upper and lower Bi2 layers are separated by large distances and thus weakly bound to the middle Ti kagome layer.

We synthesize the CTB crystal through a modified self-flux method. A typical crystal with a lateral size of over 4 mm and regular hexagonal morphology is shown in the inset of Fig. 1**b**. The corresponding x-ray diffraction (XRD) pattern confirms the pure phase and the good crystalline nature of the as-grown single crystal with a preferred [001] orientation (Fig. 1**b** and Supplementary Fig. S1). The rocking curve obtained from the (008) reflection shows a full-width-half-maximum (FWHM) of ~0.04°, along with the clear Laue diffraction spots, demonstrating the high-quality single crystal nature of the as-grown CTB (Supplementary Fig. S2). The lattice parameters *a*, *b*, and *c* are measured to be 5.839, 5.839 and 9.295 Å by single crystal diffraction, which are larger than *a* and *b* (5.548 Å), but slightly smaller than *c* (9.349 Å) of CsV$_3$Sb$_5$ crystals (Supplementary Figs. S3). The temperature-dependent XRD patterns of CTB powders prepared by grinding the single crystals to the microscale granular show indistinct structural distortion (Supplementary Fig. S4). The X-ray energy dispersive spectroscopy (EDS) performed in a scanning electron microscopy (SEM) system, determines the stoichiometric ratio of Cs:Ti:Bi =0.90:3.00:4.87 (Fig. 1**c**). The layered structures are confirmed by the atomic resolution high-angle annular dark field (HAADF) scanning transmission electron microscopy (STEM). The Z-contrast image of CTB viewed along the [210] projection is shown in Fig. 1**d**, with the structural models overlaid. The image clearly reveals the perfect crystalline structure of the CTB sample without noticeable structural defects or impurity phases. Atomic-resolution chemical analysis via electron energy-loss spectroscopy (EELS) and EDS spectrum imaging in STEM unambiguously shows clear atomic layers of Ti, Cs, and Bi atoms (Supplementary Figs. S5, S6), in excellent agreement with the single-crystal structure of CTB. These characterizations demonstrate the high-quality of the as-synthesized CTB.

Superconductivity in CTB is detected by the zero-field cooled magnetic susceptibility at various magnetic fields for *H*//c (Fig. 2**a**) and *H*//ab (supplementary Fig. S7). The onset superconducting transition temperature is ~ 4.8 K, which is significantly higher than vanadium-based kagome superconductors. The existence of the superconducting phase is unambiguously conformed by the shielding signal. A superconducting volume fraction above 60% is obtained at 1.8 K under a magnetic field of 1 μT, indicating that the superconductivity is bulk in nature. However, this superconducting phase is extremely sensitive to the applied magnetic field, whose volume fraction is suppressed down to ~10% under 0.3 mT. Superconductivity in CTB is also confirmed by electrical transport measurements. The in-plane resistivity versus temperature curve shows a metallic behavior (Fig. 2**b**) with a superconducting transition at onset $T_c$ ~ 4.8 K and zero-resistivity at ~ 3.6 K (inset of Fig. 2**b**). The scanning tunneling microscope/spectroscopy (STM/STS) measurement obtained at the surface of CTB further confirms the superconductivity. The typical d$I$/d$V$ spectrum at a temperature of 60 mK shows superconducting behaviors. The superconducting state is spatially anisotropic (Supplementary Fig. S8). In some of surface regions shows a U-shaped SC gap with a gap size of ~1.1 meV corresponding to the $T_c$~4 K, which is consistent with the electrical transport measurements mentioned above. It is noticeable that this new kagome superconductor CTB exhibits an approximately $T$-linear resistivity over a wide temperature range (~70-300 K), which might be caused by electron-phonon scatterings or electronic fluctuations. The residual resistance ratio (RRR) defined as the ratio between the resistivity at room temperature and at 5 K is about 26 (Fig. 2**b**), further supporting the relatively high quality of the as-grown CTB crystals.

In sharp contrast to vanadium-based AV$_3$Sb$_5$, the smooth resistivity versus temperature curve shows no obvious density-wave-like phase transitions. This indicates the stability of the pristine lattice structure of CTB, especially the absence of translation symmetry breaking charge density wave (CDW) formation commonly in AV$_3$Sb$_5$. We study the angular-dependent magnetoresistance (AMR) under an in-plane magnetic field of 5 T at different temperatures. As the magnetic field is rotated within the kagome plane ($\theta$=0 corresponds to magnetic field perpendicular to the current direction), we observe a pronounced two-fold rotational symmetry of the in-plane magnetoresistivity (Fig. 2**c**). Surprisingly, the magnitude of this two-fold symmetry, characterized by the AMR ratio ($\Delta\rho/\rho_{min}$=[$\rho(\theta,T)$- $\rho_{min}(T)$]/ $\rho_{min}(T)$ ×100%) (Fig. 2**d**), is extremely large, reaching up to 67% at 2 K, which is about 10 times larger than that of CsV$_3$Sb$_5$[7,14]. Although decreasing with increasing temperatures, the AMR remains at about 20% at 90 K. Besides, the shape of the two-fold anisotropy in Fig. 2**d** is remarkably elongated in the whole measured temperature range. These observations indicate an electronic nematic phase in this kagome superconductor[14].

To provide microscopic evidence for the electronic nematic state, we perform low-temperature STM/STS measurement, which is a powerful tool to study the symmetry of the electronic state with sub-lattice spatial resolution[15-20]. The large-sized STM topography, $T(E, \mathbf{r})$, shows a clean Bi surface (see Method and Supplementary Fig. S9) with a few types of randomly-distributed point defects (*e.g.* at $E$=40 mV in Fig. 3**a**). The quasiparticle interference (QPI) features around point defects are weak to be observed in the STM topography. Only six Bragg peaks of nearly equal intensity consistent with the crystal symmetry are visible in the power spectral density of the Fourier transform (FT) of STM topography $T(E, \mathbf{q})$ (*e.g.* at $E$=40 mV in Fig. 3**b**). In contrast, the QPI around the point defects in the d$I$/d$V$ maps, $g(E, \mathbf{r})$ (*e.g.* $E$=40 mV in Fig. 3**c**), is much stronger and its FT, $g(E, \mathbf{q})$, shows rich QPI patterns (*e.g.* at $E$=40 mV in Fig. 3**d**). Around the zone center Γ point, there is a small hexagon, a larger hexagon rotated 30 degrees, and an even larger fragmented circle delineated by six arcs. The intensities of these QPI features have approximated six-fold ($C_6$) symmetry. Intriguingly, a $C_2$ symmetric petal-like pattern is clearly seen in $g(E, \mathbf{q})$ in Fig. 3**d**, whose intensity along the direction Γ-K1 is much more stronger than at the corresponding locations in the other two directions (Γ-K2 and Γ-K3). The QPI patterns are robust and reproducible with different STM tips scanning over an identical region (see supplementary Fig. S10). Since the $C_2$ symmetric features are absent in topography, but present in the d$I$/d$V$ maps on selective contours rather than in all the QPI patterns, we attribute their origin to a genuine nematic electronic phase instead of structural distortions. Note that although rotational symmetry is broken at the onset of the CDW transition in AV$_3$Sb$_5$[6,10,19,21], the electronic state in the CDW phase is not nematic because of the simultaneous broken translation symmetry. Moreover, with decreasing temperatures well into the superconducting state, the unidirectional QPI patterns are observed as well (Supplementary Fig. S11), demonstrating that the kagome CTB is a nematic superconductor.

In order to gain microscopic insight for the absence of the CDW order and the emergence of electronic nematic order, we perform first-principle density functional theory (DFT) calculations of the electronic structure for the CTB crystal (details given in supplementary Fig. S12). While the band structure carries a nontrivial topological $Z_2$ invariant analogous to the vanadium-based kagome metals AV$_3$Sb$_5$[2,8], the low energy orbital-resolved band dispersion (Fig. 4**a** and Supplementary Fig. S12) reveals substantial differences. For of all, the strength of the SOC is much stronger due to the heavier Bi atom and causes a

~ 400 meV downward shift of the Bi $p_z$ band near the zone center and large splitting of the Dirac crossings at the K points. Remarkably, the Fermi level is pushed below two van Hove singularities (VHS) of the in-plane $d_{xy}/d_{x^2-y^2}$ orbitals at M points (highlighted by the light-yellow shade in Fig. 4a), signifying dramatically altered carrier density of the "135" transition-metal d-orbitals for broadened physical landscape of the kagome metals. Moreover, the out-of-plane $d_{xz}$, $d_{yz}$ dominated bands are significantly modified near the M points compared to AV$_3$Sb$_5$ due to the large and hybridization and SOC, such that the corresponding VHS are removed. These important changes in the electronic structure intertwined with the stable phonon spectrum[22,23] provide a microscopic origin for the stability of the pristine kagome lattice in CsTi$_3$Bi$_5$ and the absence of the $2a_0 \times 2a_0$ CDW order observed in all AV$_3$Sb$_5$.

There are four bands crossing the Fermi level determined by the DFT with similar dispersions in the $k_z=0$ and $k_z=\pi$ planes (Fig. 4a). Correspondingly, the constant-energy contours (CEC) near the Fermi level consist of four sheets with different orbital contents as shown in Fig. 4b. The blue circle around the Γ/A point corresponds to the electron-like band (Fig. 4a) originating from the Bi-$p_z$ orbital. The six small green circles around the M/L points are derived from the Ti $d_{z^2}$ orbital. The cyan colored inner hexagon and the six triangular petals around the K/H points are dominated by the Ti out-of-plane $d_{xz}$ and $d_{yz}$ orbitals, while the outer red hexagon comes predominately from the Ti in-plane $d_{xy}$ and $d_{x^2-y^2}$ orbitals. The QPI patterns imaged by the STS conductance map $g(V, q)$ can thus be compared directly to the joint density of states calculated for the electronic states on the CEC. Figures 4c and 4d illustrate such a comparison between $g(20\ meV, q)$ and the simulated QPI from the calculated band states of CTB at $k_z=0$ planes (calculations show that the energy band is only weakly $q_z$ dependent) in Fig. 4b. The detected QPI patterns mainly originate from five dominating scattering branches ($q_{1,2,3,4,5}$) marked in Fig. 4b: intra-band scattering of the Bi $p_z$ orbital ($q_1$), inter-band scattering between the Bi $p_z$ orbital and the Ti $d_{xy}/d_{x^2-y^2}$ orbitals ($q_2$), intra-band scattering of the $d_{xz}$, $d_{yz}$ orbitals ($q_3$), intra-band scattering of $d_{xy}/d_{x^2-y^2}$ orbitals ($q_4$), and intra-band scattering of Ti-$d_{z^2}$ orbital ($q_5$). As a result, we can identify that only the $q_3$ branch corresponding to the intra-band quasiparticle scattering involving the Ti–$d_{xz}$ and $d_{yz}$ orbitals (the triangular pedals in Fig. 4c) breaks the rotational-symmetry, indicating that the electronic nematicity is orbital-selective.

To further confirm the orbital-selectivity of the nematic order, and determine the energy scale over which the electronic states break rotation symmetry, we collected the energy-dependent d$I$/d$V$ maps g($E$,**r**) and the corresponding FT image g($E$,**q**). We find that the $C_2$-symmetryic features in g($E$,**q**) get suppressed at the sample bias below ~-120 mV and above ~280 mV (details see Supplementary Fig. S13, four exemplary results are shown in Fig. 4e). In the cut of g($E$,**q**), the dispersion of four branches **q**$_1$, **q**$_2$, **q**$_3$ and **q**$_4$ are obtained. The **q**$_1$, **q**$_2$ and **q**$_4$ branches show electron-like dispersions with distinct Fermi vectors while **q**$_3$ exhibits a hole-like dispersion (Fig. 4f). To study the symmetry-breaking feature of each branch, we then compare the coherent quasiparticle weights of three branches along two equivalent ***q*** directions, Γ-K1 and Γ-K3 (Fig. 4f). The **q**$_1$, **q**$_2$ and **q**$_4$ branches show the same weight along the two ***q*** directions, while the **q**$_3$ branch shows stronger weight along Γ-K1 direction than the one along Γ-K3 direction.

The combined results of the measured QPI and the DFT calculations convincingly demonstrate that the lowering of the rotational symmetry in CsTi$_3$Bi$_5$ is due to nematic order predominantly involving the out-of-plane Ti–$d_{xz}$ and $d_{yz}$ orbitals. The significant energy range over which nematic electronic structure is observed suggests that the onset of nematic order may occur at relatively high temperatures. To uncover the origin of these novel phenomena requires the understanding of the multiorbital correlation effects in the "135" kagome metals, which is currently lacking, and may stimulate further experimental and theoretical studies. Our observations are consistent with a plausible mechanism for lowering the D$_{6h}$ crystalline symmetry through rotational symmetry breaking hybridizations among $d_{xz}$ and $d_{yz}$ orbitals with the atomic onsite D$_{2h}$ symmetry. Interestingly, such intra and inter-orbital bond nematic orders can arise from the extended (inter-site) Coulomb interaction[24], which have been studied in connection to the nematic phase of the tetragonal iron-based superconductor FeSe[25].

In summary, we have successfully fabricated high quality crystals of Ti-based kagome metal CTB and discovered its bulk superconductivity ($T_c$ ~ 4.8K) for the first time. Through a combination of angular-dependent magnetoresistance and STM/STS spectroscopic imaging, supplemented by first-principle DFT calculations, we find that the electronic states of CTB lower the crystalline symmetry by purely rotational-symmetry breaking and exhibit an orbital-selective nematic phase. The Ti-based "135" kagome metals thus provide a new materials platform for the much-needed exploration of the multi-orbital correlation effects, orbital or bond ordered electron liquid crystal phases, and the implications on the helical Dirac fermion topological surface states and bulk superconductivity.

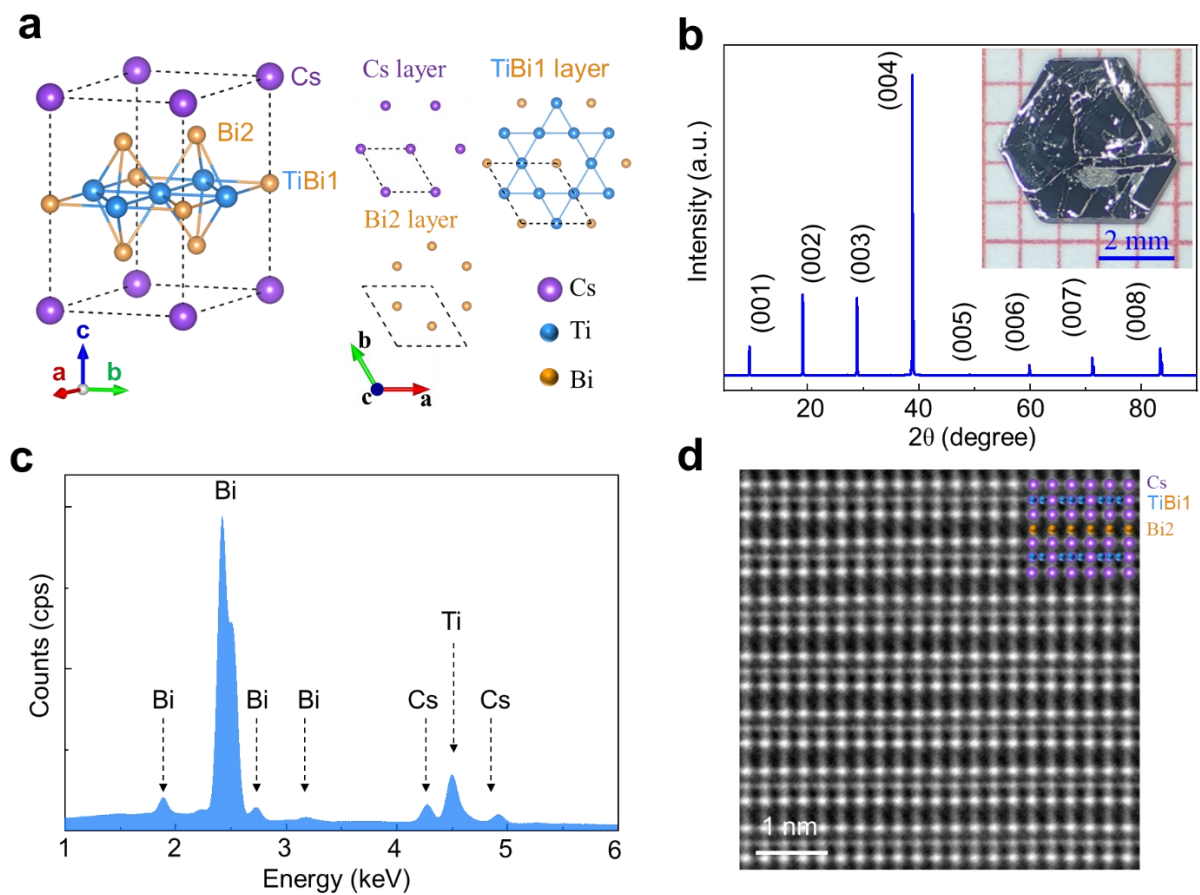

**Fig. 1. Atomic structure and chemical analysis of the Ti-based kagome superconductor. a.** Schematic of the structure of $CsTi_3Bi_5$ crystal with Cs atoms in purple, Bi atoms in light orange, Ti atoms in azure. The dashed lines represent a unit cell. The perfect kagome net of Ti atoms is mixed with a simple triangular net of Bi1 atoms. **b**, XRD pattern of the as-prepared $CsTi_3Bi_5$ single crystals with the corresponding Miller indices (00$l$), showing the high quality of the crystal. Inset is a photograph of the as-prepared $CsTi_3Bi_5$ single crystal with a regular hexagonal morphology and sharp edges. The size of the crystal is over 4 mm. **c,** EDS of the $CsTi_3Bi_5$ crystal, showing the stoichiometric ratio of Cs:Ti: Bi =0.90 : 3.00 : 4.87. **d,** Atomic resolution STEM-ADF Z-contrast image of $CsTi_3Bi_5$ viewed along the [210] projection, with the atomic structural models overlaid.

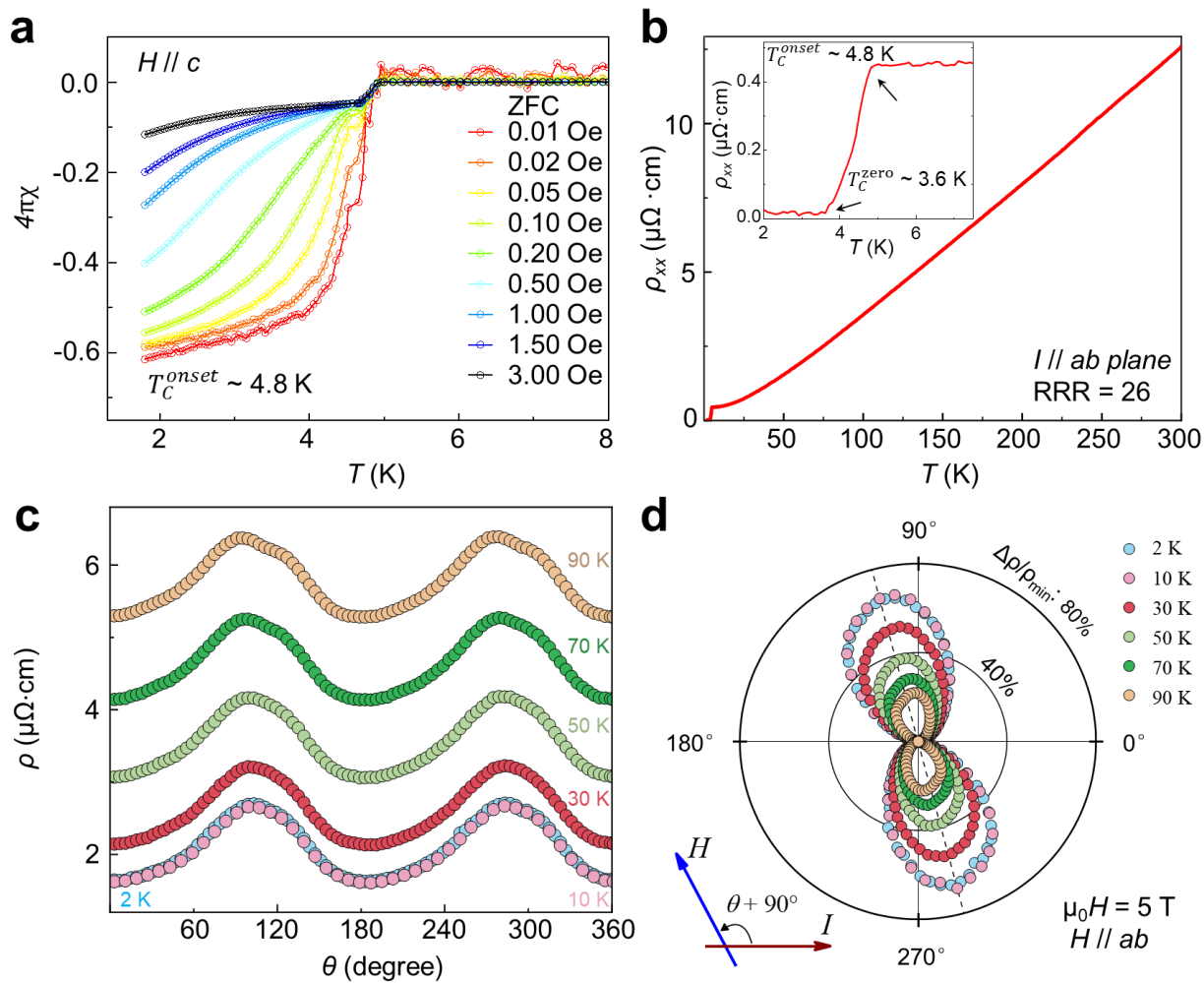

**Fig. 2. Characterization of superconductivity and two-fold rotational symmetry. a**, The magnetic susceptibilities under various magnetic fields for $H // c$ between 2 K and 8 K, showing the magnetic shielding effect. **b**, Temperature-dependent resistivity between 2 K and 300 K under zero magnetic field. The inset shows the superconducting transition with $T_c^{onset} \sim 4.8$ K and $T_c^{zero} \sim 3.6$ K. **c**, The in-plane magnetoresistivity as a function of $\theta$ at different temperatures. **d**, Polar plots of AMR = $\Delta\rho/\rho_{min}$ = [$\rho(\theta, T)$ -$\rho_{min}(T)$]/$\rho_{min}(T) \times 100\%$ from (c). **c** and **d** show the two-fold rotational symmetry in the AMR remarkably. $\theta$ in (**c, d**) is the angle between the directions of the external field ($\mu_0 H$ = 5 T) and the current ($I$), and $\theta$ = 90º corresponds to $H // I$. The $\theta$ is illustrated by the schematic diagram.

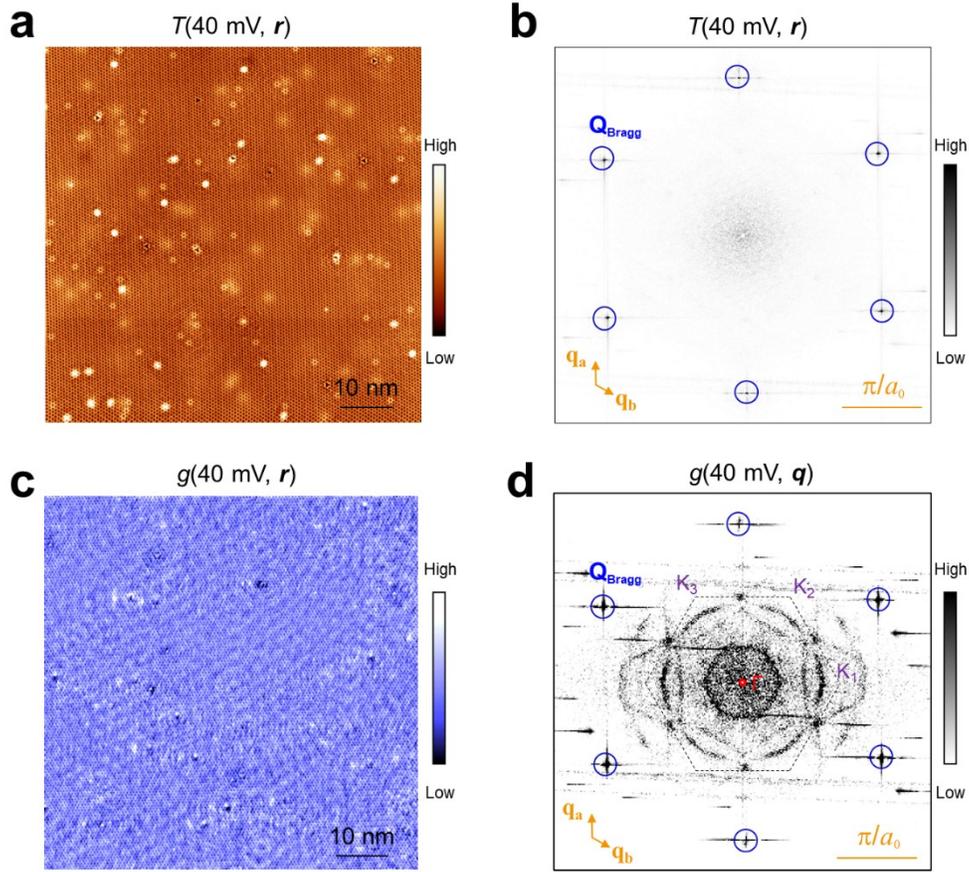

**Fig. 3. STM topography, d$I$/d$V$ maps and corresponding Fourier transform of Bi surface in CsTi$_3$Bi$_5$ crystal. a,b** The STM topography of Bi surface $T$(40 mV, $r$) and the magnitude of drift-corrected Fourier transform (FT) $T$(40 mV,$q$), showing isotropic six Bragg peaks ($V_s$=-10 mV, $I_t$=500 pA ). **c, d,** d$I$/d$V$ map $g$(40 mV, $r$) and the magnitude of drift-corrected FT $g$(40 mV, $q$), revealing the C$_2$ symmetric QPI patterns. The flower-like QPI patterns along Γ-K1 direction highlighted by the red dotted line show stronger intensity than those along Γ-K2 and Γ-K3 directions. The C$_2$ symmetric QPI patterns is only observed in the d$I$/d$V$ maps indicate an electronic nematicity at the Bi surface and 4.2 K.

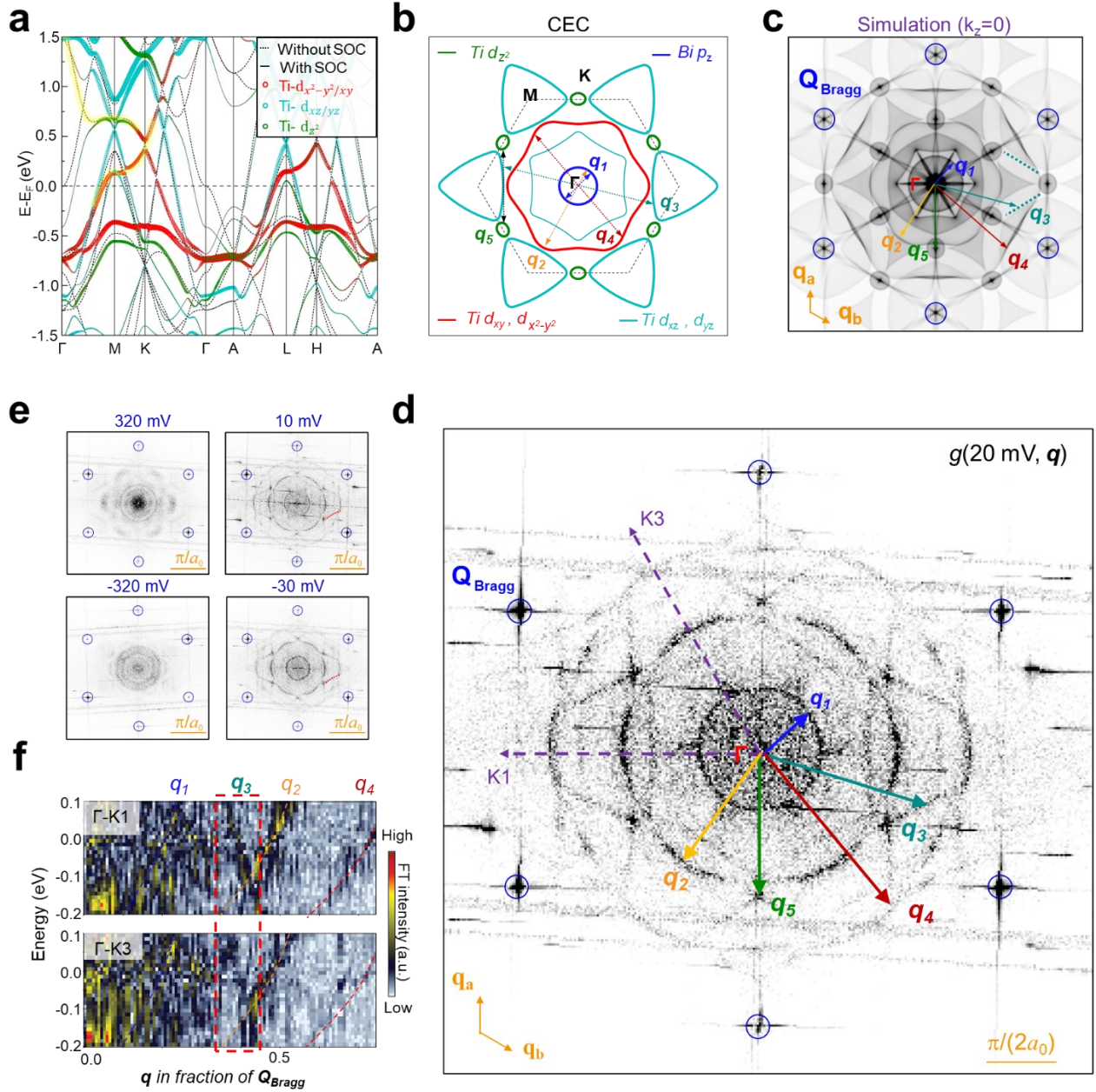

**Fig. 4. Orbital-selectivity nematicity in CsTi$_3$Bi$_5$. a**, The electronic energy bands calculated with SOC (red solid curves) and without SOC (blue dotted curves). The contribution from different Ti orbitals are highlighted by distinct colors. **b,** Calculated CEC around zero energy and k$_z$=0, showing four pockets contributed from distinct orbital bands as labelled. Five scattering vectors correspond to the **q$_1$**, **q$_2$**, **q$_3$**, **q$_4$** and **q$_5$**. The consistence between QPI measurements and simulations demonstrates that the C$_2$ symmetric

features originate from $q_3$ branch corresponding to the Ti–$d_{xz}$, $d_{yz}$ orbitals. **c,d** The QPI simulations (**c**) based on calculated band structures and $g(20\ \text{mV},q)$ (**d**), show similar QPI patterns consisting of vectors $q_1$, $q_2$, $q_3$, $q_4$ and $q_5$ ($V_s$=-40 mV, $I_t$=500 pA, $V_{mod}$=5 mV). **e,** $g(E,q)$ at $E$=-320 meV, -30 mV, 10 mV and 320 meV, respectively, showing that the $C_2$-symmetric QPI patterns disappear at the relatively-large energy beyond the Fermi surface. **f,** Comparison of FT cut along Γ-K1 and Γ-K3 direction, respectively, showing that the intensity of brunch corresponding to $q_3$ scattering vectors along Γ-K1 is stronger than the one along Γ-K3 (highlighted by the dotted red line), further supporting that the nematicity originates from $q_3$ branch corresponding to the Ti–$d_{xz}$, $d_{yz}$ orbitals.